\documentclass[12pt]{article}
\usepackage{amsfonts}
\usepackage{amsmath}
\usepackage{amssymb}
\usepackage{graphicx}
\usepackage{float}
\usepackage{color}
\textwidth=6.5in \hoffset=-.75in \voffset=-1in \numberwithin{equation}{section}
\numberwithin{figure}{section}

\newcommand {\nn}{\nonumber}
\newcommand {\be}{\begin{equation}}
\newcommand {\ee}{\end{equation}}
\begin{document}

\begin{titlepage}
\vspace{1cm}
\begin{center}
{\Large \bf {Atiyah-Hitchin in Five Dimensional Einstein-Gauss-Bonnet Gravity}}\\
\end{center}
\vspace{1cm}
\begin{center}
{M. Butler\footnote{mdb815@mail.usask.ca}$^*$, A. M. Ghezelbash\footnote{masoud.ghezelbash@usask.ca}$^*$, E. Massaeli\footnote{erfan.massaeli@gmail.com}$^\dagger$, M. Motaharfar\footnote{mmotaharfar2000@gmail.com}$^\dagger$}
\\
$^*$Department of Physics and Engineering Physics,  University of Saskatchewan, \\
Saskatoon, SK S7N 5E2, Canada\\
$^\dagger$Department of Physics, Shahid Beheshti University, Evin,
Tehran 19839, Iran
\vspace{1cm}
\vspace{2cm}
\end{center}

\begin{abstract}

We construct a new class of stationary exact solutions to five-dimensional Einstein-Gauss-Bonnet gravity. The solutions are based on four-dimensional self-dual Atiyah-Hitchin geometry. We find analytical solutions to the five-dimensional metric function that are regular everywhere. We find some constraints on the possible physical solutions by investigating the solutions numerically. We also study the behavior of the solutions in the extremal limits of the Atiyah-Hitchin geometry. In the extremal limits, the Atiyah-Hitchin metric reduces to a bolt structure and Euclidean Taub-NUT space, respectively. In these limits, the five-dimensional metric function approaches to a constant value and infinity, respectively. We find the asymptotic metrics are regular everywhere.

\end{abstract}
\end{titlepage}\onecolumn 
\bigskip 

\section{Introduction}

In the low-energy limit, the M-theory describes effectively the eleven dimensional supergravity \cite{NEW1, NEW2, NEW3}. Hence, the brane solutions in supergravity furnish classical soliton states of M-theory which in turn motivate considerable interest in finding the supersymmetric brane solutions \cite{NEW4, NEW5}.
New supergravity solutions for localized D2/D6, D2/D4, NS5/D6 and NS5/D5 intersecting brane systems were discovered recently on transverse Bianchi type IX space \cite{GH}. One special feature of the solutions is that the solutions do not need to be in the near core region.  These solutions generalize the other known M2- and M5- branes in the transverse  Taub-NUT and Taub-Bolt, Eguchi-Hanson and Atiyah-Hitchin backgrounds \cite{Cherkis}, \cite{GH2, GH3}. 

The Atiyah-Hitchin geometry is the crucial non-trivial part to the moduli space of two monopole solutions of Bogomolnyi equation \cite{KS, KS2}.  The Atiyah-Hitchin geometry belongs to the set of four-dimensional self-dual curvature geometries. The self-dual property in four-dimensional geometry is a result of hyper-Kahlericity of the moduli space of two monopole solutions \cite{MEAH}. One special feature of Atiyah-Hitchin geometry is that the metric is given entirely in terms of three functions on the monopoles separation. The hyper-Kahlericity of the geometry implies the three functions satisfy a set of ordinary coupled first-order differential equations. The Atiyah-Hitchin geometry has been recently used to construct  the non-stationary exact cosmological solutions to five-dimensional Einstein-Maxwell-Chern-Simons theory with positive cosmological constant \cite{MEAH} as well as exact solutions to the five dimensional Einstein-Maxwell Theory \cite{MEAH2}. Moreover, in \cite{HAR}, the authors found the small corrections to the asymptotic limit of Atiyah-Hitchin manifold with the correct topology at infinity that corresponds to the bound states of instantons and anti-instantons. Moreover, the various generalizations of Atiyah-Hitchin space are identified with the full quantum moduli space of $N$ = 4 supersymmetric gauge theories in three dimensions \cite{GAH}. The intriguing character of Atiyah-Hitchin geometry is that the metric tensor depends on three independent functions of distance between the two monopoles in the Bogomolnyi solution.  The three metric functions are the solutions to 
a Darboux-Halpern system that guarantees the Atiyah-Hitchin space is a self-dual curvature geometry. 

Inspired with the recent interest in finding the exact solutions to generalization of Einstein gravity with higher order corrections to Einstein-Hilbert action \cite{GB1}-\cite{GB8} and new exact convoluted solutions \cite{GH4}-\cite{GH7}, in this paper, we construct new exact solutions to Gauss-Bonnet gravity in five dimensions where the spatial section of the solution is Atiyah-Hitchin geometry.

The paper is organized as follows. In section \ref{sec2}, we review briefly the five-dimensional Einstein-Gauss-Bonnet theory and the field equations as well as the Atiyah-Hitchin space and its features. In section \ref{sec3},
we choose a proper choice for one of the Atiyah-Hitchin metric functions and then analytically solve three field equations of the Einstein-Gauss-Bonnet theory to find the behavior of the metric function. We then numerically verify that other non-zero field equations for the Einstein-Gauss-Bonnet theory are indeed satisfied.  We find and present numerical solutions for the metric function in five dimensions and discuss the behavior of the physical metric functions. In section \ref{sec4}, we consider the extremal limits of the Atiyah-Hitchin metric and discuss the properties of the five-dimensional solutions. In the extremal limits, the Atiyah-Hitchin metric reduces to a bolt structure and Euclidean Taub-NUT space, respectively. In these limits, the five-dimensional metric function approaches to a constant value and infinity, respectively.

\section{Five-dimensional Einstein-Gauss-Bonnet gravity based on Atiyah-Hitchin geometry}
\label{sec2}
The Einstein-Gauss-Bonnet action in five dimensions is given by \cite{GB}-\cite{GB3}
\begin{equation}
S= \frac{1}{16 \pi \text{G}} \int d^5 x \sqrt{-g} \left(R + \alpha \mathcal{L}_{GB} \right),
\label{eqn:EGB_action}
\end{equation}
where $g$ is the determinant of metric, $G$ is Newton's Gravitational constant, $R$ is Ricci scalar, $\alpha$ is the Gauss-Bonnet parameter. In (\ref{eqn:EGB_action}),  the term $\mathcal{L}_{GB}$ is the Gauss-Bonnet term that is quadratic in Riemann tensor and is given by
\begin{equation}
\mathcal{L}_{GB} \equiv  R_{abcd}R^{abcd} - 4R_{ab}R^{ab} + R^2,
\end{equation}
where $R$, $R_{ab} $ and $R_{abcd}$ are Ricci scalar, Ricci tensor and Riemann tensor respectively.  The variation of the action (\ref{eqn:EGB_action}) with respect to the metric tensor yields the following gravitational field equations 
\begin{equation}
{ \bf {\cal R}}_{\mu\nu}=0,
\end{equation}
where
\begin{eqnarray}
 { \bf {\cal R}}_{\mu\nu}&=&R_{\mu \nu} - \frac{1}{2}Rg_{\mu \nu}\nn\\
 &-&
\alpha \left\{ \ \frac{1}{2}g_{\mu \nu}(R_{abcd}R^{abcd} - 4R_{ab}R^{ab} + R^2) - 2RR_{\mu \nu}+4R_{\mu a}R^{a}_{\nu} + 4R^{a b}R_{\mu a \nu b} - 2R^{abc}_{\mu}R_{\nu a b c}\right\}.\notag\\
\label{eqn:EGB_fieldequations}
\end{eqnarray}
Inspired with the results in \cite{MEAH} and \cite{MECQG}, we consider the following ansatz for the five dimensional metric such as
\begin{equation}
ds_5^{2}=-H(r)^{-2}dt^{2}+H(r)ds_{AH}^2,
\label{ds5}
\end{equation}
where $ds_{AH}^{2}$ represents the four dimensional Atiyah-Hitchin geometry. The metric for the Atiyah-Hitchin space is given by the following $SO(3)$ invariant form 
\begin{equation}
ds_{AH}^{2}=f(r)^{2}dr^{2}+a(r)^{2}\sigma _{1}^{2}+b(r)^{2}\sigma
_{2}^{2}+c(r)^{2}\sigma _{3}^{2},  \label{AHmetric}
\end{equation}
where $\sigma_i,\, i=1,2,3$ represent the Maurer-Cartan one-forms
\begin{eqnarray}
\sigma _{1}&=&-\sin \psi d\theta +\cos \psi \sin \theta d\phi, \label{mcFORMS1}\\ 
\sigma _{2}&=&\cos \psi d\theta +\sin \psi \sin \theta d\phi, \label{mcFORMS2}\\ 
\sigma _{3}&=&d\psi +\cos \theta d\phi. \label{mcFORMS3}
\end{eqnarray}
We note that  $\sigma _{i\text{ }}$ satisfy the following property 
\begin{equation}
d\sigma _{i}=\frac{1}{2}\varepsilon _{ijk}\sigma _{j}\wedge \sigma _{k}.
\label{dsigma}
\end{equation}
The self-duality of the Atiyah-Hitchin geometry (\ref{AHmetric}) as well as four-dimensional Einstein's equations imply that the metric functions $a(r)$, $b(r)$ and $c(r)$ satisfy the first order coupled differential equations
\begin{eqnarray}
\frac{da}{dr}&=&f\frac{(b-c)^{2}-a^{2}}{2bc}, \label{conditions1}\\ 
\frac{db}{dr}&=&f\frac{(c-a)^{2}-b^{2}}{2ca}, \label{conditions2}\\ 
\frac{dc}{dr}&=&f\frac{(a-b)^{2}-c^{2}}{2ab}. \label{conditions3}
\end{eqnarray}
The metric ansatz (\ref{ds5}) leads to the following three coupled non-linear differential equations (\ref{eqn:EGB_fieldequations}) for the metric function $H(r)$ and its first and second derivatives that are given by
\begin{eqnarray}
\label{rr}
\notag
{ \bf {\cal R}}_{rr}&=&\frac{1}{4 f^{2}  a^{3} b^{3} c^{3} H^{5}} \Bigg(\bigg\{ 6\, a^{3} b^{3} c^{3} (H')^{4}
+12\,a^{2}  b^{2} c^{2} \Big((  b' c + c' b  ) a + a' c b  \Big) H(H')^{3}
\\\notag&&+24\, a b c  \bigg( a b c ( c\,b' +  b\,c') a'
+ a^{2} b c \,b'  c' +\frac{1}{12}\, f^{2}( a +b+c)( a +b-c ) ( a -b -c  )  ( a-b +c  )\bigg) H^{2}  (H')^{2}
\\\notag&&+48\, \bigg[ -\frac{a  f^{2} }{12}\, \Big( c \,b'\big(  a^{4}+2(  b^{2}- c^{2} )  a^{2}-3\, b^{4}+2\, b^{2} c^{2}+ c^{4} \big)  +b\, c'\big(  a^{4}-2( b^{2} -c^{2})  a^{2}+ b^{4}+2\, b^{2} c^{2}-3\, c^{4} \big)   \Big)
\\\notag&& +b c\, a' \Big(a^{2}b c \,b' c' +\frac14\, f^{2} \big( a^{4}-\frac23 ( b^{2}+ c^{2})  a^{2}-\frac13\, ( b-c)^{2}(b+c) ^{2} \big)  \Big)  \bigg]  H^{3}H'\bigg\}\alpha
\\\notag&& +4\,a b c f^2  \bigg( a b c  (c\,b' + b\,c' ) a' +  a^{2}b c \,b' c'+\frac{1}{4}\, f^{2}( a +b+c)( a +b-c ) ( a -b -c  )  ( a-b +c  )\bigg) H^5
\\&&-3\,  a^{2} b^{2} c^{2}H^3  (H')^{2}\Bigg),
\end{eqnarray}
\begin{eqnarray}
\label{thph}
\notag
{ \bf {\cal R}}_{\theta\phi}&=&-\frac{1}{ f^{5} a^{2} b^{2} c^{2} H^{5}}\Bigg(\bigg\{ 3\, H   f   \bigg(  a^{2} b^{2} c^{2}(  a^{2}- b^{2})(H')^{2}-\frac43\, a b c \Big(  a b (-a^{2}+ b^{3}) c'+c( b^{3}a'-a^{3}b')  \Big)H\,H'
\\\notag&&+ \Big( -\frac43\, a   b   c   ( b^{3}a'-a^{3}b')c'
+ f^{2}(a^2 - b^2)\big(  a^{4}+ \frac23(  b^{2}- c^{2} )  a^{2}+ b^{4}-\frac23\, b^{2} c^{2}-\frac13\, c^{4} \big)\Big)  H^{2}\bigg)  H'' 
\\\notag&&+  f a b c \Big(  2\,a b( a^{2}-b^{2})(H')^{2}-4 (b^{3}a'- a^{3}b' )H\, H' \Big) c'' H^{2}  - f   a   b^{4} c   \Big( 4\,  c'   H  H'  +2\, c    (H')^{2} \Big)  a''  H^{2}
\\\notag&&  +  f   a^{4} b   c   \Big( 4\,  c'   H  H'  +2\, c    (H')^{2} \Big)  b'' H^{2}   -\frac92\, f   a^{2} b^{2} c^{2} (a^2- b^2)(H')^{4}+6\, a b c\Big(  \frac12 a b c( -a^{2}+ b^{2})f'
\\\notag&& +\big(( - a^{3} b+ a b^{3})c'+ c( b^{3}a'-  a^{3}b')\big)f   \Big)H (H')^{3}
-\frac{15}2\,\bigg( -\frac45\,a b c \Big( a b ( - a^{2}  +  b^{2} )  c'  + c   (      b^{3}a'-     a^{3}b' )  \Big)f'  
\\\notag&& -\frac45\, f a b c (  b^{3}a'- a^{3}b')c' + f^{3}(a^2  - b^2)\Big(a^{4}+ \frac23( b^{2}- c^{2}) a^{2}+ b^{4}-\frac23\, b^{2} c^{2}-\frac13\, c^{4} \Big)   \bigg)  H^{2}   (H')^{2}
\\\notag&&-3\,\Big( -4\, a   b   c  (a' b^{3}- b' a^{3})c'  +f^{2}(a^{2}- b^{2}) (a^{4}+ ( \frac23\, b^{2}-\frac23\, c^{2} )  a^{2}+ b^{4}-\frac23\, b^{2} c^{2}-\frac13\, c^{4} )\Bigg)f' H^{3}H'   \bigg\}\alpha
\\\notag&&+  f^{3} a^{2} c   b^{2}(a^2-b^2) c''  H^{5} - f^{3} a   b^{4} c^{2} a''  H^{5} +f^{3} a^{4} b   c^{2}  b''  H^{5}  +\frac34\, f^{3} a^{2} b^{2} c^{2}(a^2-b^2)   H^{3} (H')^{2}
\\\notag&&+  f^{2} \bigg(  a b c\Big(  a b( - a^{2}+b^{2} )  c'  + c (b^{3}a'- a^{3}b')\Big)f'+ f\Big(- a b c(b^{3}a'-a^{3}b')c'+f^{2}(a^2- b^2)\big( \frac34\, a^{4}
\\&&+ \frac12(  b^{2}-c^{2} )  a^{2}+\frac34\, b^{4}-\frac12\, b^{2} c^{2}-\frac14\, c^{4} \big)  \Big) \bigg) H^{5}\Bigg)\sin ( \theta ) \sin ( \psi ) \cos ( \psi ),
\end{eqnarray}
\nopagebreak

and

\begin{eqnarray}
\label{psps}
\notag
{ \bf {\cal R}}_{\psi\psi}&=&\frac1{ 4\, f^{5} a^{2} b^{2}H^{5}}\Bigg(\bigg\{  16\, f   \bigg( \frac34\, a^{2} b^{2} c^{2} ( H')^{2}+ a   b   c^{2} (  b  \, a'  +a\,b')H\, H'+\Big(a b c^{2}\, a' b'
\\\notag&&-\frac14\, f^{2} \big(a^{4}+ 2(- b^{2}+c^{2})  a^{2}+ b^{4}+2\, b^{2} c^{2}-3\, c^{4}\big)  \Big)H^{2}\bigg)  H\,  H''+16\,f a b c^{2} \Big( b' H\,H'+\frac12\,b(H')^{2}\Big)a'' H^{2}  
\\\notag&&+16\, f a  b   c^{2} \Big(   a'    H'   H  +\frac12\,  (H')^{2} a\Big)  b'' H^{2}  -18\, f   a^{2} b^{2} c^{2}  (H')^{4}-24\,a b   c^{2}\Big( \frac12\,a b\,f'  + f (  b a'  + a b' )\Big)H\, (H')^{3}
\\\notag&&-24\,  \Big(  a   b   c^{2} (  b   a'  + a   b'   )f'  + f a b c^{2}\,a' b'-\frac {5}{12}\, f^{3} \big(  a^{4}+ 2( - b^{2}+c^{2} )  a^{2}+ b^{4}+2\, b^{2} c^{2}-3\, c^{4} \big)    \Big) H^{2} (H')^{2}
\\\notag&&-48\, \bigg( a b c^{2}\,a' b'-\frac1{12}\, f^{2} \Big(  a^{4}+ 2( -b^{2}+c^{2} )  a^{2}+ b^{4}+2\, b^{2} c^{2}-3\, c^{4} \Big)\bigg)   f' H^{3} H' \bigg\}\alpha 
\\\notag&&+ 4\,f^{3} a b^{2} c^{2}  a''  H^{5}  +4\,f^{3} a^2  b   c^{2} b''  H^{5}  +3\,f^{3} a^{2} b^{2} c^{2} H^{5}(H')^{2}+4\,  f^{2} \bigg(-a b c^{2}( b\, a'+ a\, b') f'
\\&&+ f \Big(a   b   c^{2}a' b'+  \frac14 f^{2}\big( - a^{4}+ 2( b^{2}- c^{2} )  a^{2}- b^{4}-2\, b^{2} c^{2}+3\, c^{4}\big)  \Big)  \bigg) H^{5} \Bigg),
\end{eqnarray}
where the prime and double prime denote $\frac{{\rm d}}{{\rm d}r}$ and $\frac{{\rm d^2}}{{{\rm d}r}^2}$, respectively. In section \ref{sec3}, we analytically solve these three equations to find the explicit exact solutions for the metric function $H(r)$. We note that there are four more non-zero coupled non-linear differential equations, i.e. ${ \bf {\cal R}}_{tt},{ \bf {\cal R}}_{\theta\theta},{ \bf {\cal R}}_{\phi\phi},{ \bf {\cal R}}_{\psi\phi}$. We show numerically in next section that all the field equations are indeed satisfied in the limit of small Gauss-Bonnet parameter.

\section{Exact solutions to the field equations}
\label{sec3}
Inspired with the results in \cite{MEAH}, we consider the Atiyah-Hitchin metric function $f(r)$ as
$f(r)=4\,a(r)b(r)c(r)$, hence we get the following form for the Atiyah-Hitchin metric (\ref{AHmetric}) as 
\begin{equation}
ds_{AH}^{2}=16\,a(r)^{2}b(r)^{2}c(r)^{2}dr^{2}+a(r)^{2}\sigma _{1}^{2}+b(r)^{2}\sigma
_{2}^{2}+c(r)^{2}\sigma _{3}^{2}. \label{AHmetric2}
\end{equation}
The functions $a(r)$, $b(r)$ and $c(r)$ satisfy equations (\ref{conditions1}--\ref{conditions3}) which upon transformations
\begin{eqnarray}
a(r)^{2}&=&\frac{\psi_2(r)\psi_3(r)}{4\,\psi_1(r)}\label{psis1},\\
b(r)^{2}&=&\frac{\psi_3(r)\psi_1(r)}{4\,\psi_2(r)}\label{psis2},\\
c(r)^{2}&=&\frac{\psi_1(r)\psi_2(r)}{4\,\psi_3(r)}\label{psis3},
\end{eqnarray}
yield the following Darboux-Halpern differential system for the functions $\psi_i(r),\,i=1,2,3$
\begin{eqnarray}
\frac{d}{dr}\Big(\psi_1(r)+\psi_2(r)\Big)+2\,\psi_1(r)\psi_2(r)&=&0,\label{H1}\\
\frac{d}{dr}\Big(\psi_2(r)+\psi_3(r)\Big)+2\,\psi_2(r)\psi_3(r)&=&0,\label{H2}\\
\frac{d}{dr}\Big(\psi_3(r)+\psi_1(r)\Big)+2\,\psi_3(r)\psi_1(r)&=&0.\label{H3}
\end{eqnarray} 
The Darboux-Halpern differential system (\ref{H1}--\ref{H3}) has the following solutions
\begin{eqnarray}
\psi_1(\vartheta)&=&-\frac{1}{2}\Big(\frac{d}{d\vartheta}\mu^2+\frac{\mu^2}{\sin\vartheta}\Big),
\label{psi1} \\
\psi_2(\vartheta)&=&-\frac{1}{2}\Big(\frac{d}{d\vartheta}\mu^2-\frac{\mu^2\cos\vartheta }{\sin\vartheta}\Big),
\label{psi2} \\
\psi_3(\vartheta)&=&-\frac{1}{2}\Big(\frac{d}{d\vartheta}\mu^2-\frac{\mu^2}{\sin\vartheta}\Big),
\label{psi3} 
\end{eqnarray}
where 
\begin{equation}
\mu (\vartheta)=\frac{1}{\pi}\sqrt{\sin\vartheta}K(\sin\frac{\vartheta}{2}). \label{varw}
\end{equation}

\begin{figure}[H]
\centering
\includegraphics[scale=0.30]{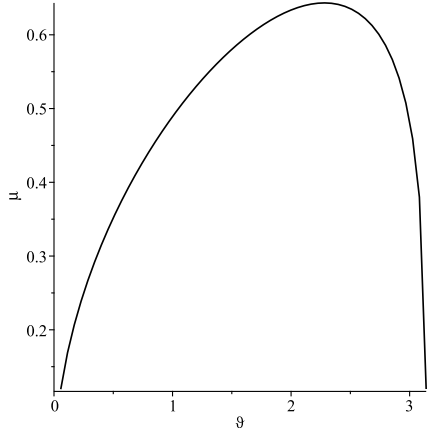}
\caption{The function $\mu$ versus  $\vartheta$, as given by equation (\ref{varw}).} 
\label{mufig}
\end{figure}

The new coordinate $\vartheta$ is related to the coordinate $r$ by 
\begin{equation}
r=-\int _\vartheta ^ \pi \frac{d\vartheta}{\mu(\vartheta)^{2}},
\label{theta}
\end{equation}
which has a monotonic increasing behaviour as a function of $r$.  We note that the range of the new coordinate $\vartheta$ is over $[0, \pi]$, where we choose the range of the coordinate $r\in (-\infty ,0]$. 

Figure \ref{mufig} shows the dependence of $\mu$ to $\vartheta$ as given by equation (\ref{varw}). We notice from figure (\ref{mufig}) that the function $\mu (\vartheta)$ has an increasing behavior from $\vartheta=0$ to $\vartheta_0=2.281$. At 
$\vartheta=\vartheta_0$, the
function $\mu$ reaches to maximum value $0.643$ and decreases then to zero at $\vartheta=\pi$. In conclusion, in the range of $0 < \vartheta <\pi$, the function $\mu$ is positive definite and so
the transformation (\ref{theta}), is completely well defined. In figure \ref{PSIN}, we present numerical solutions to the Darboux-Halpern differential equations (\ref{H1})-(\ref{H3}) that are given in (\ref{psi1})-(\ref{psi3}).

\begin{figure}[H]\centering
	\includegraphics[scale=0.30]{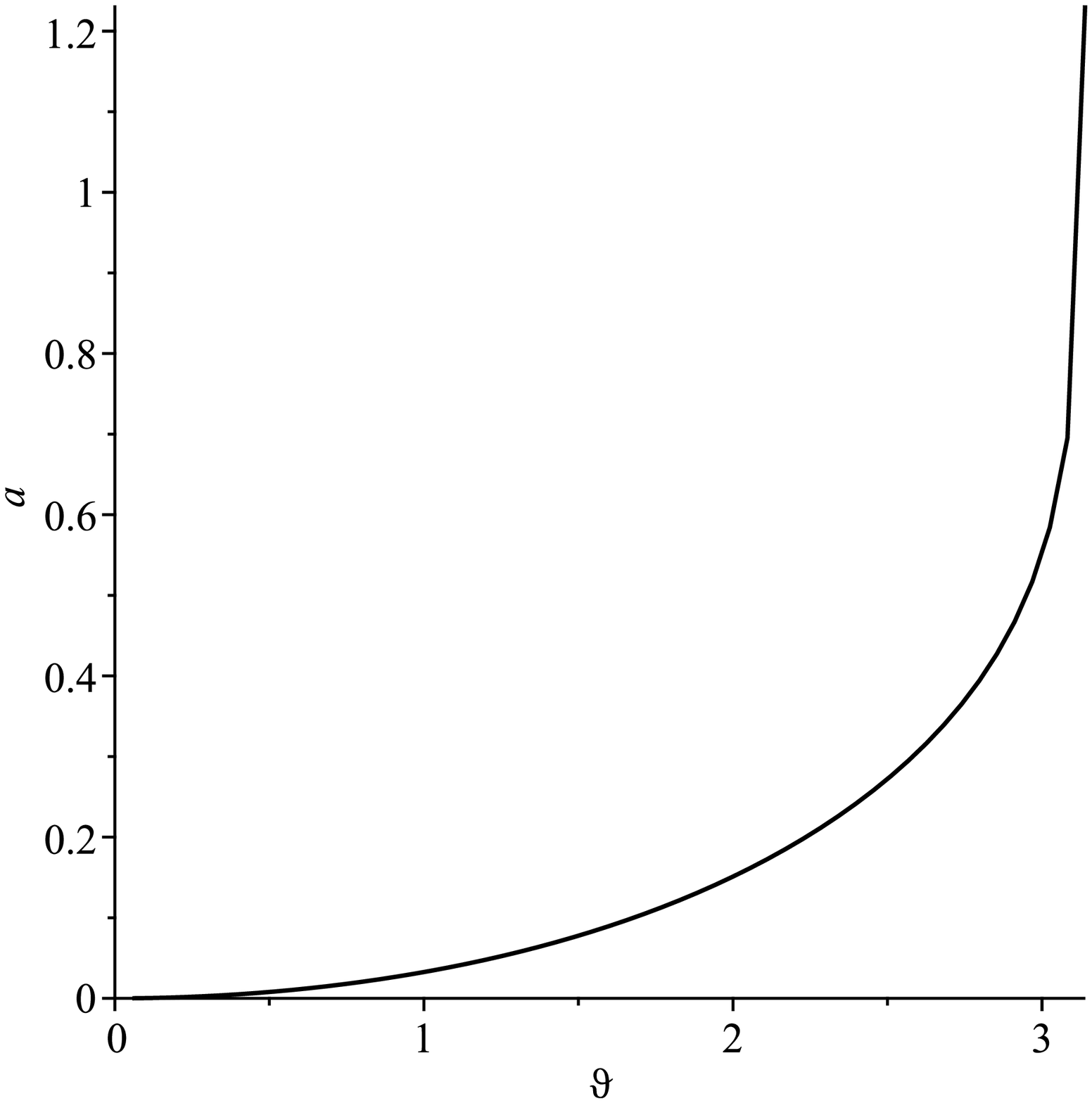} \includegraphics[scale=0.30]{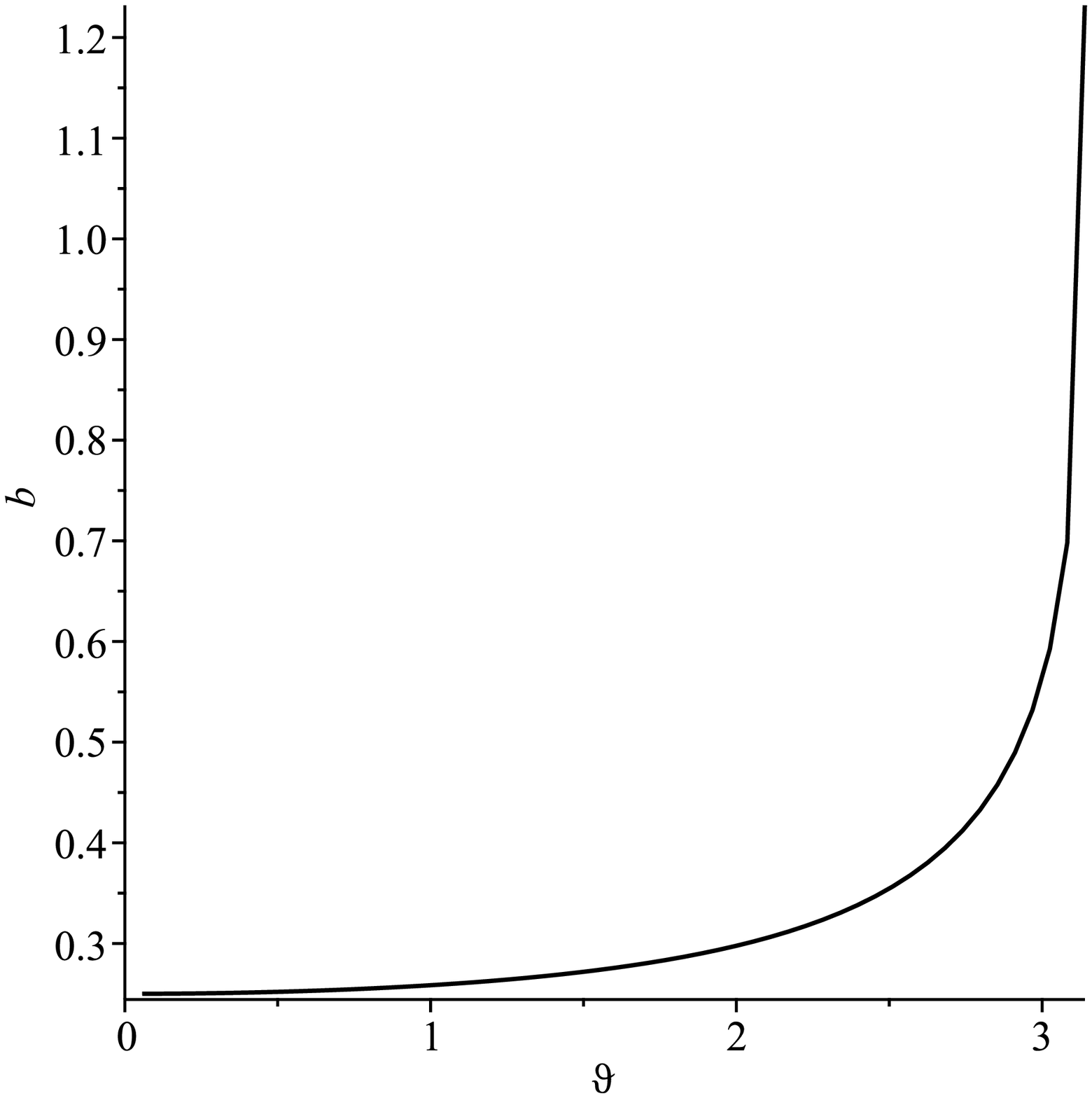}\\
	\centering
	\includegraphics[scale=0.30]{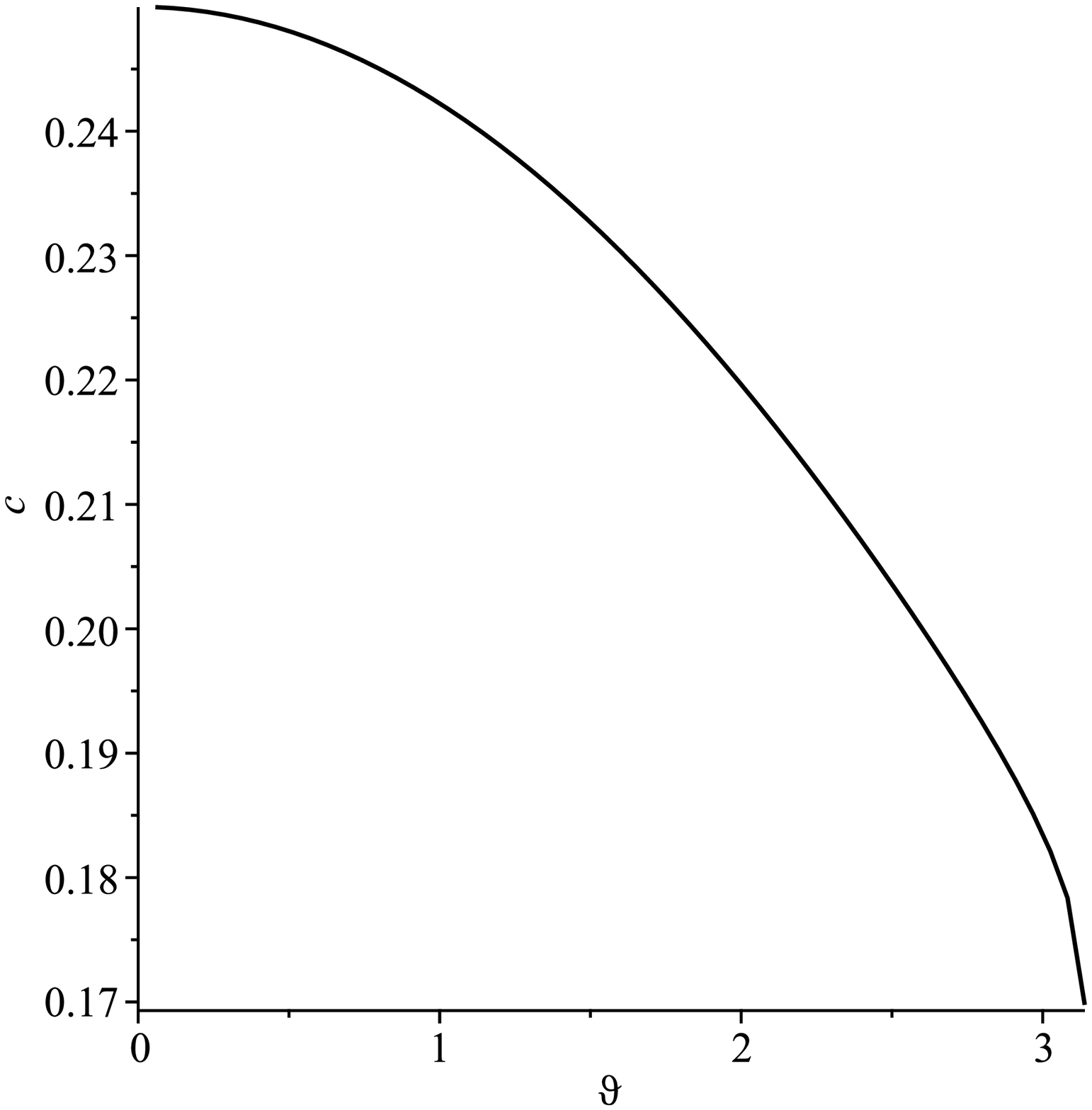}
	\caption{Numerical Solutions for  the Atiyah-Hitchin metric functions $a(\vartheta)$, $b(\vartheta)$, and $c(\vartheta)$.} 
	\label{abcN}
\end{figure}

\begin{figure}[H]\centering
	\includegraphics[scale=0.30]{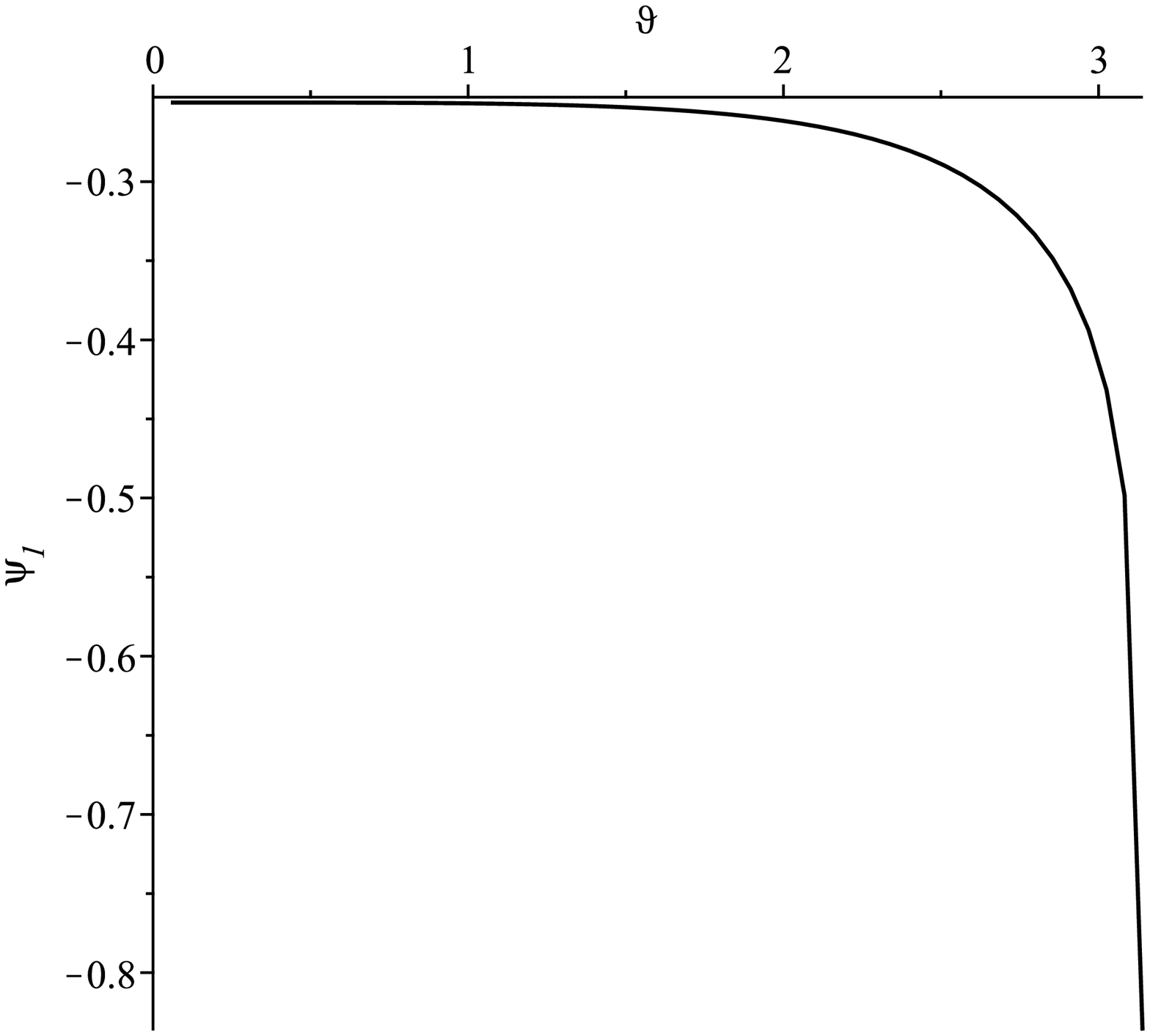} \includegraphics[scale=0.30]{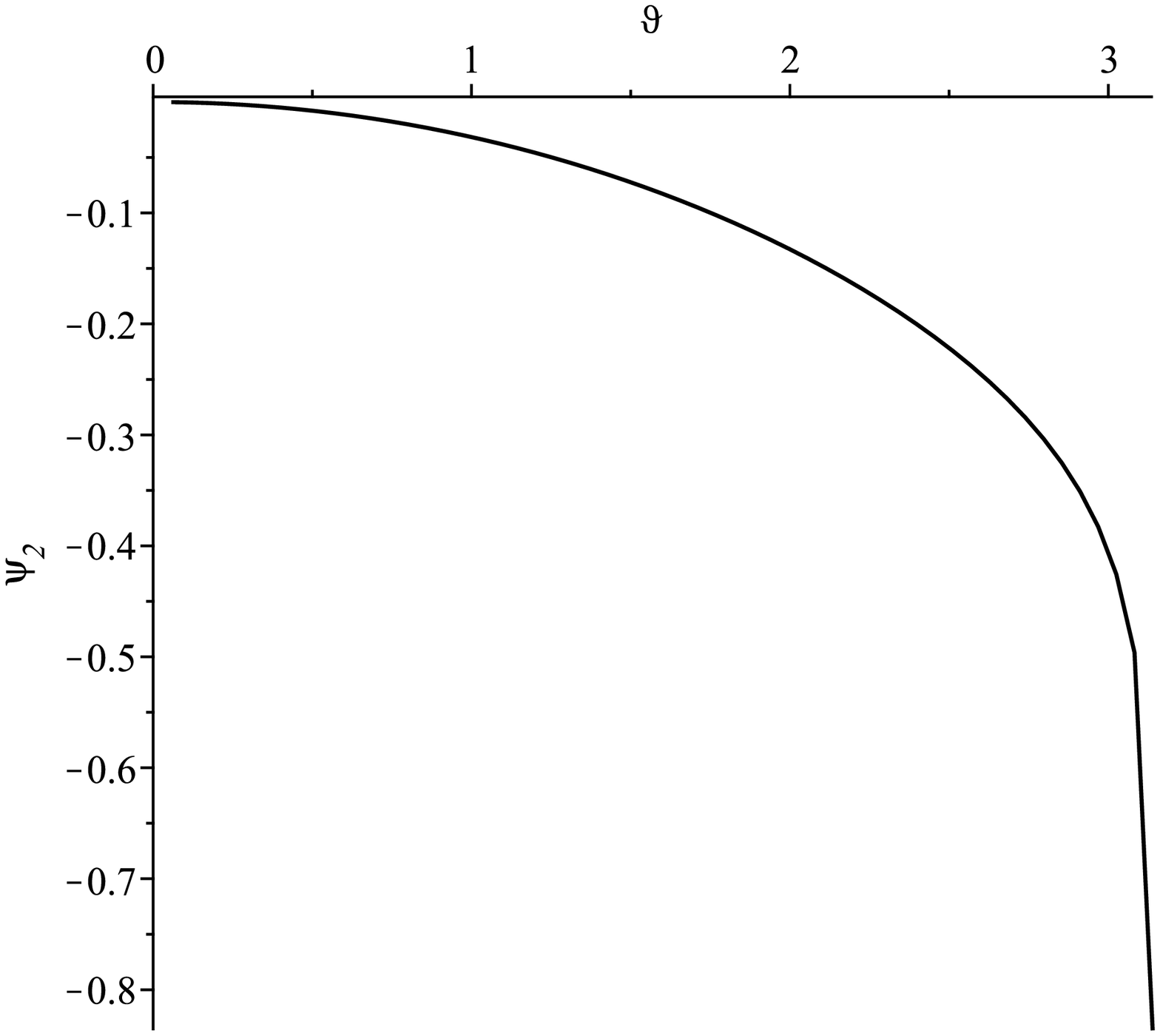}\\
	\centering
	\includegraphics[scale=0.30]{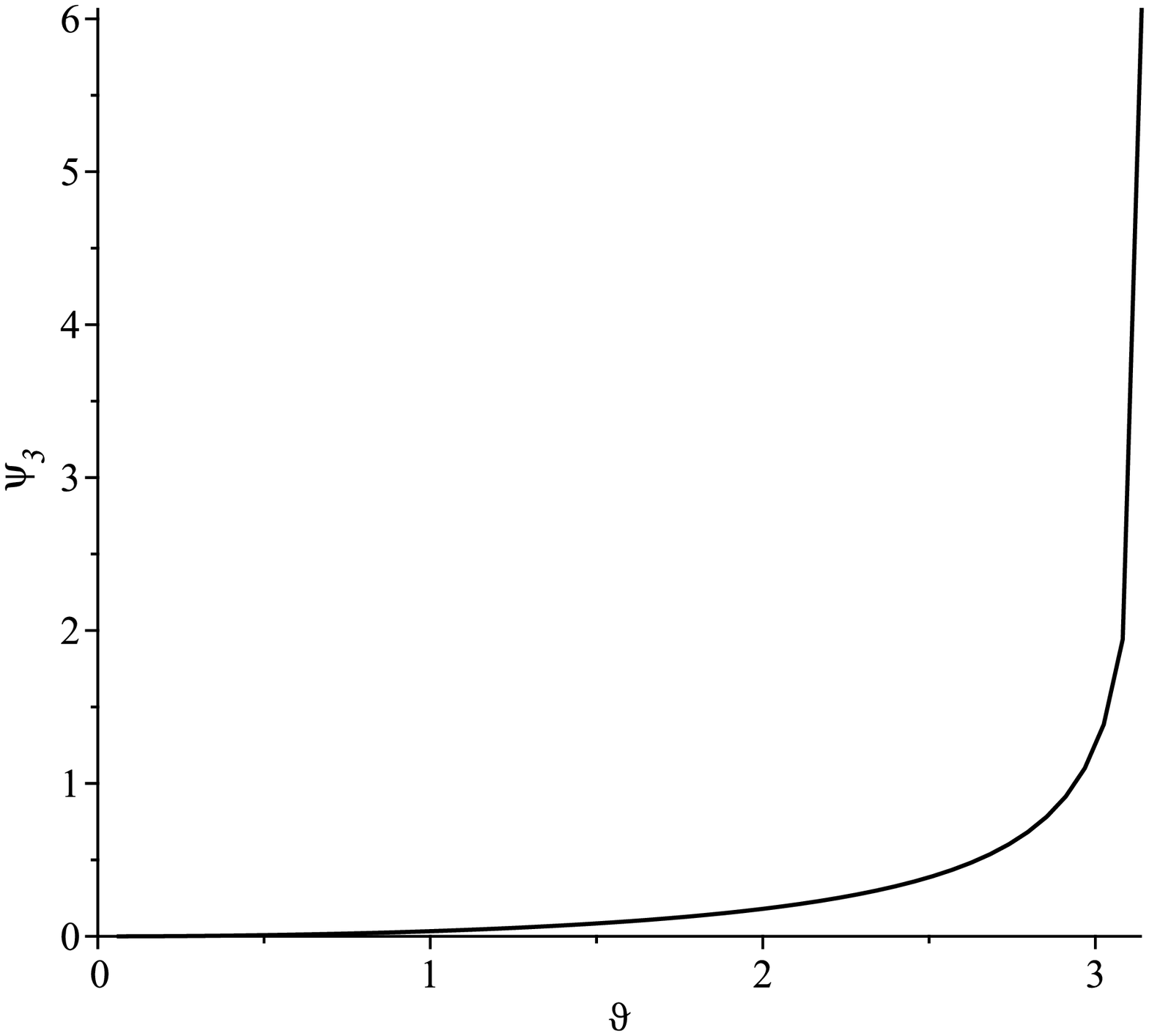}
	\caption{Numerical Solutions for Darboux-Halpern functions $\psi_1$, $\psi_2$, and $\psi_3$ versus $\vartheta$.} 
	\label{PSIN}
\end{figure}

As it is obvious from figure \ref{PSIN}, the Darboux-Halpern functions $\psi_1,\psi_2$ are negative definite, while $\psi_3$ is positive definite. This shows that the right hand sides of equations (\ref{psis1}), (\ref{psis2}) and (\ref{psis3}) are indeed positive definite.
We find the solutions for the Atiyah-Hitchin metric functions $a(\vartheta)$, $b(\vartheta)$ and $c(\vartheta)$ are given explicitly by 
\begin{eqnarray}
a(\vartheta)&=&\frac{1}{2\pi}\sqrt{\Big( E(\sin\frac{\vartheta}{2}) -(\cos\frac{\vartheta}{2})^{2}  K(\sin \frac{\vartheta}{2})\Big)\Big(K(\sin\frac{\vartheta}{2}) - E(\sin\frac{\vartheta}{2})\Big){K(\sin\frac{\vartheta}{2})}/{E(\sin\frac{\vartheta}{2})}},\nn\\ &&\label{a-th}
\\b(\vartheta)&=&\frac{1}{2\pi}\sqrt{\Big( K(\sin\frac{\vartheta}{2}) -  E(\sin \frac{\vartheta}{2})\Big){K(\sin\frac{\vartheta}{2})}\times{E(\sin\frac{\vartheta}{2})}/\Big(E(\sin\frac{\vartheta}{2}) - (\cos\frac{\vartheta}{2})^{2}K(\sin\frac{\vartheta}{2})\Big)},\nn\\ &&\label{b-th}
\\ c(\vartheta)&=&\frac{1}{2\pi}\sqrt{\Big( E(\sin\frac{\vartheta}{2}) -(\cos\frac{\vartheta}{2})^{2}  K(\sin \frac{\vartheta}{2})\Big){K(\sin\frac{\vartheta}{2})}\times{E(\sin\frac{\vartheta}{2})}/\Big(K(\sin\frac{\vartheta}{2}) - E(\sin\frac{\vartheta}{2})\Big)},\nn\\ &&\label{c-th}
\end{eqnarray}
where $K$ and $E$ are the complete elliptic integrals
\begin{eqnarray}
K(k) &=&\int_{0}^{1}\frac{dt}{\sqrt{1-t^{2}}\sqrt{1-k^{2}t^{2}}}%
=\int_{0}^{\pi /2}\frac{d\theta }{\sqrt{1-k^{2}\cos ^{2}\theta }},
\label{Ell} \\
E(k) &=&\int_{0}^{1}\frac{\sqrt{1-k^{2}t^{2}}dt}{\sqrt{1-t^{2}}}%
=\int_{0}^{\pi /2}\sqrt{1-k^{2}\cos ^{2}\theta }d\theta.
\end{eqnarray}
In figure \ref{abcN}, we present numerical solutions for the Atiyah-Hitchin metric functions $a,b$, $c$ versus $\vartheta$. 
We consider now the new quantities $ ~x=\frac{{\rm d Ln}(H)}{{\rm d}r},~~ y=\frac{{\rm d^2 Ln}(H)}{{{\rm d}r}^2}+\left(\frac{{\rm d Ln}(H)}{{\rm d}r}\right)^2$ and re-write the field equations (\ref{rr})--(\ref{psps}) as
\begin{eqnarray}
&&\alpha\Big(a_1 x^4+a_2 x^3+a_3 x^2+a_4 x\Big)-a_5 x^2 H=0,
\label{rr2}
\\&&\alpha\Big(\big(2 a_1 c_1 x^2+c_2 x+c_3\big) y-3 a_1 c_1 x^4 +c_4 x^3+c_5 x^2+c_6 x \Big)+a_5 c_1 x^2 H=0,
\label{thph2}
\\&&\alpha\Big(\big(2 a_1 x^2+ b_1 x +b_2\big) y-3 a_1 x^4+ b_3 x^3+b_4 x^2+b_5 x\Big)+a_5 x^2 H=0.
\label{psps2}
\end{eqnarray}

From equations (\ref{thph2}) and (\ref{psps2}), we find that $y$ is given in the term of $x$, by 
\begin{eqnarray}
y=\frac{\bigg((c_4-c_1 b_3)x^3+(c_5-c_1 b_4)x^2+(c_6-c_1 b_5)x\bigg)}{(c_2-c_1 b_1)x+c_3-c_1 b_1}.
\end{eqnarray}
We then find a quartric equation for $x$ from equations (\ref{rr2}) and (\ref{thph2}) as
\begin{eqnarray}
d_1 x^4+d_2 x^3+d_3 x^2+ d_4 x+d_5 =0,\label{QUA}
\end{eqnarray}
where the coefficients $d_i,\, i=1,\cdots ,5$ are completely in terms of Atiyah-Hitchin metric functions $a$, $b$ and $c$. In Appendix A, we present all the coefficients $d_i,\, i=1,\cdots ,5$ explicitly. The equation (\ref{QUA}) has 4 solutions that we call them $x_i,\,i=1,\cdots ,4$, repectively.  We present the explicit form of $x_i,\,i=1,\cdots ,4$ in terms of Atiyah-Hitchin metric functions $a$, $b$ and $c$  in Appendix B. 
From equation (\ref{rr2}), we find the corresponding metric functions $H_i,\,i=1,\cdots , 4$ as
\begin{eqnarray}
H_i=\alpha\frac{(a_1 x_i^3+a_2 x_i^2+a_3 x_i+a_4 )}{a_5 x_i }.
\end{eqnarray}
Of course, we can express the four exact solutions $H_i$ explicitly in the term of $\vartheta$ by using (\ref{a-th})--(\ref{c-th}), however the expression are extremely long and so we do not present them here.  Plugging the solutions $H_i$ in four remaining field equations ${ \bf {\cal R}}_{tt},{ \bf {\cal R}}_{\theta\theta},{ \bf {\cal R}}_{\phi\phi},{ \bf {\cal R}}_{\psi\phi}$ yields very long expressions that is almost impossible to verify analytically that they satisfy the field equations. As a result, we switch to numerical methods, to show that the four field equations are indeed satisfied by numerical calculation.  

\begin{figure}[H]
\centering
	\includegraphics[scale=0.30]{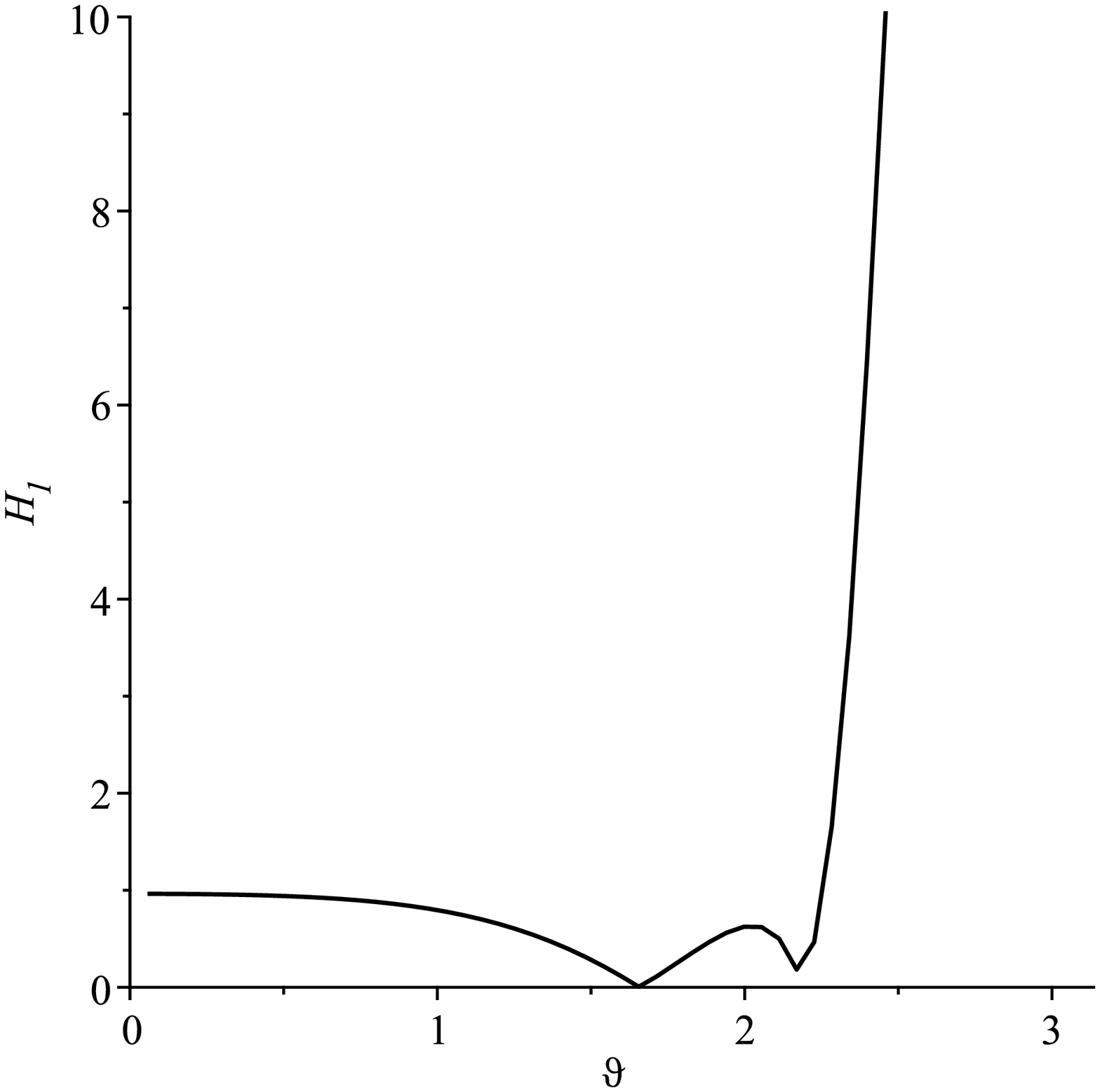} 
	\includegraphics[scale=0.30]{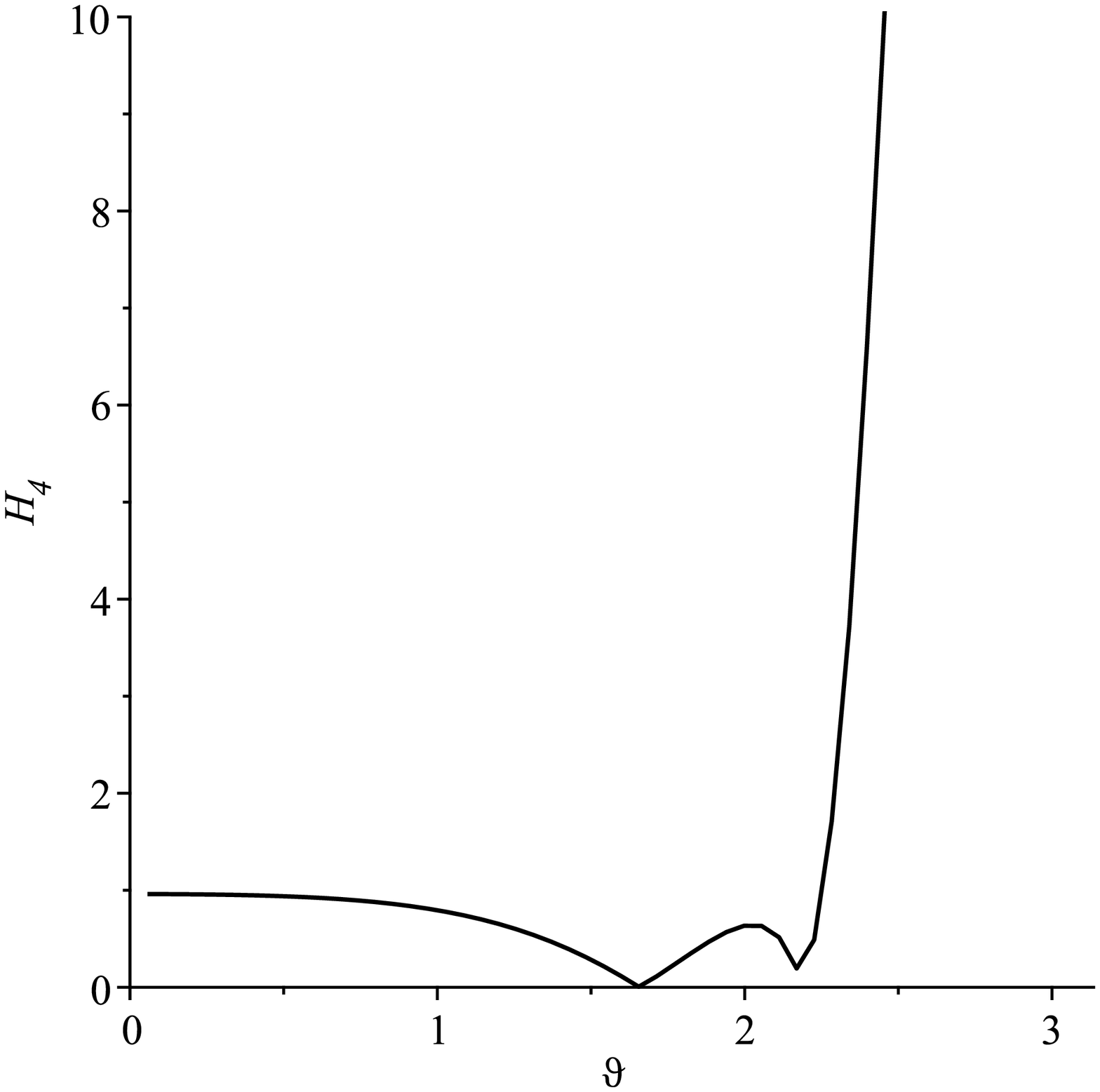}\\
	\caption{Numerical solutions for the scaled metric functions $H_1$ and $H_4$ versus $\vartheta$.} 
	\label{HN}
\end{figure}

\begin{figure}[H]
\centering
	\includegraphics[scale=0.30]{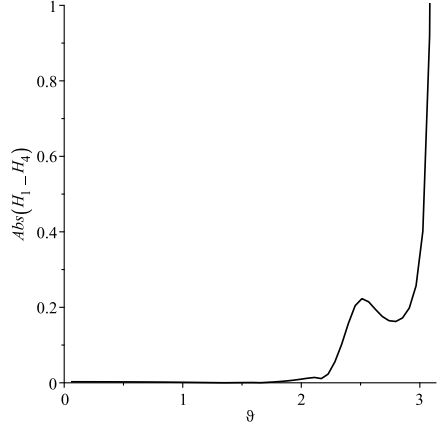} 
	\caption{The scaled absolute difference between metric functions $H_1$ and $H_4$.} 
	\label{diffig}
\end{figure}

Our numerical approach begins using the equations (\ref{psi1})-(\ref{psi3}) that are solutions to the Darboux-Halpern differential system \eqref{H1}-\eqref{H3}. Since we can numerically approximate the elliptical integral $K$, we can construct numerical solutions for \eqref{varw}.  With the solutions for $\mu(\vartheta)$, we can produce approximations for $\psi_1$, $\psi_2$, $\psi_3$ in \eqref{psi1}-\eqref{psi3} and by extension, solutions for $a(\vartheta)$, $b(\vartheta)$, $c(\vartheta)$ with equations \eqref{psis1}-\eqref{psis3}. 
From these we can reduce the original field equations for a given value of $\vartheta$ into a manageable form. Utilizing the independent reduced field equations, we can rearrange for $H''$, $H'$ and substitute and solve for $H$ algebraically, an approach that was intractable with the full equations. Using our systems of coefficients in Appendices A and B, we are able to map our approximations to the analytical solutions for $H_1$, $H_2$, $H_3$, and $H_4$.  Plugging the numerical solutions $H_i,\,i=1,\cdots ,4$ in four field equations ${ \bf {\cal R}}_{tt},{ \bf {\cal R}}_{\theta\theta},{ \bf {\cal R}}_{\phi\phi},{ \bf {\cal R}}_{\psi\phi}$ reveals that they are indeed satisfy all the field equations, up to the numerical values of less or equal to $10^{-41}$. The numerical solutions for $H_2$ and $H_3$ are not positive everywhere for $ 0  \leq \vartheta \leq \pi$, hence they are not physical solutions for the metric function in (\ref{ds5}). In figure \ref{HN}, we present the results of numerical calculations for the metric functions $H_1$ and $H_4$ for $\alpha=0.1$.
We note that the vertical axis is scaled by a factor of $10^{-13}$. Both $H_1$ and $H_4$ show a decreasing behavior from $\vartheta=0$ to around $\vartheta=1.6554$ where $H_1$ reaches to a local minimum of $5.9401 \times 10^{10}$ and $H_4$ reaches to a minimum of $6.1562 \times 10^{10}$. The metric functions $H_1$ and $H_4$, then increase to $6.2450 \times 10^{12}$ and $6.3408 \times 10^{12}$ respectively at $\vartheta=1.9979$. They again decrease to local minima of $1.8306 \times 10^{12}$ and $1.9419 \times 10^{12}$ respectively at $\vartheta=2.1691$. After that, they increase monotonically as $\vartheta \rightarrow \pi$. Although from figure \ref{HN}, it seems that $H_1$ is very similar to $H_4$, however we should note that for all values of $\vartheta$ between $0$ and $1.3699$, $H_1 > H_4$, while for $\vartheta$ between $1.3699$ and $\pi$, $H_1 < H_4$.  In figure \ref{diffig}, we plot the scaled absolute value of difference between $H_1$ and $H_4$ for $\vartheta$ between $0$ and $\pi$ where the vertical axis is scaled by a factor of $10^{-13}$. Figure \ref{diffig} shows that $H_1$ and $H_4$ are indeed two independent metric functions for the five-dimensional spacetime \eqref{ds5}.

\section{Extremal limits of the solutions}
\label{sec4}

In this section, we investigate the extremal limits of the metric \eqref{ds5} where the coordinate $r$ approaches the extremal limit $-\infty$ (that corresponds to $\vartheta=0$) as well as $r\rightarrow 0$ (that corresponds to $\vartheta=\pi$). In the limit of $\vartheta\rightarrow 0$, the Atiyah-Hitchin metric functions \eqref{a-th} -\eqref{c-th} become

\begin{eqnarray}
a(\vartheta)&=&\frac{\vartheta ^2}{768}(24+\vartheta^2+O(\vartheta^4)),   \label{aatxiinf}\\
b(\vartheta)&=&\frac{1}{4}(1+\frac{\vartheta ^2}{32}+O(\vartheta^4)),   \label{batxiinf}\\
c(\vartheta)&=&\frac{1}{4}(1-\frac{\vartheta ^2}{32}+O(\vartheta^4)),  \label{catxiinf}
\end{eqnarray}
respectively. Moreover, we find that the function $\mu$ in \eqref{varw}, in the limit of $\vartheta\rightarrow 0$, has the power series expansion as
\be 
\mu = \frac{1}{2}\vartheta ^{1/2} (1+\frac{1}{16}\vartheta^2+O(\vartheta^4)),
\ee
and so from equation \eqref{theta}, we find that in the limit of $\vartheta\rightarrow 0$, 
\be
r = 4 \ln \vartheta+O(\vartheta^2).
\ee
We find the asymptotic form for the Atiyah-Hitchin metric \eqref{AHmetric2} as
\be
ds_{AH{\vert}{\vartheta\rightarrow 0}}^2 =\frac{1}{4}d\varepsilon^2+\varepsilon ^2 \sigma_1^2+\frac{1}{16}(\sigma_2^2+\sigma_3^2),
\ee
where $\varepsilon=\frac{1}{32}\vartheta ^2$.
Hence, we find the following asymptotic for the five-dimensional metric \eqref{ds5}
\be
ds_5=-H_0^{-2}dt^2+H_0(\frac{1}{4}d\varepsilon^2+\varepsilon ^2 \sigma_1^2+\frac{1}{16}(\sigma_2^2+\sigma_3^2)),\label{5as1}
\ee
where $H_0$ is the value of metric function $H_1$ or $H_4$ as $\vartheta \rightarrow 0$, according to figure \ref{HN}. The metric \eqref{5as1} is quite regular everywhere. In fact, the
Ricci scalar of the metric \eqref{5as1} is equal to 
\be
R=-\,32\,\frac{(2\varepsilon -1)(2\varepsilon+1)}{H_0},
\ee
while the Kretschman invariant is given by
\be
{\cal K}=1024\,\frac{176\varepsilon^4-24\varepsilon^2+13}{H_0^2}. 
\ee
In the other extremal limit, where $\vartheta \rightarrow \pi$, we find that the Atiyah-Hitchin metric function $a(\vartheta)$ (equation \eqref{a-th})  becomes equal to $b(\vartheta)$ (equation \eqref{b-th}). In fact, we find $a=b= \frac{-1}{2\pi}\ln (\frac{\pi-\vartheta}{8})$, where $\vartheta \rightarrow \pi$. In the extremal limit $\vartheta \rightarrow \pi$, the other Atiyah-Hitchin metric function $c(\vartheta)$ approaches to constant number $\frac{1}{2\pi}$. Hence in the extremal limit $\vartheta \rightarrow \pi$, the Atiyah-Hitchin metric \eqref{AHmetric2} becomes 
\be
ds_{AH{\vert}{\vartheta\rightarrow \pi}}^2 =\frac{1}{4\pi^2}(d{\cal N}^2+{\cal N}^2d\Omega^2+(d\psi+\cos\theta d\phi)^2),
\label{AHTN}
\ee
where ${\cal N}=-\ln(\pi-\vartheta)$ and $d\Omega ^2$ is the metric on unit sphere, parametrized by $(\theta,\phi)$.
We note that the asymptotic metric \eqref{AHTN} is the Euclidean Taub-NUT geometry \cite{TNbib}. The asymptotic geometry for the five-dimensional metric \eqref{ds5} is 
\be
ds_{5}^2 =-H_\infty^{-2}dt^2+\frac{H_\infty}{4\pi^2}(d{\cal N}^2+{\cal N}^2d\Omega^2+(d\psi+\cos\theta d\phi)^2),\label{5as2}
\ee
where $H_\infty$ is the diverging value of the metric function $H_1$ or $H_4$ as $\vartheta \rightarrow \pi$, according to figure 3.4. The metric \eqref{5as2} is also regular everywhere.  We find that the Ricci scalar and the Kretschman invariant of the metric \eqref{5as2} are given by
\be
R=-\,\frac{2\pi^2}{H_\infty {\cal N}^4},
\ee
and
\be
{\cal K}=4\,\pi ^4\frac{48{\cal N}^2+11}{H_\infty^2{\cal N}^8},
\ee
respectively.

\section{Concluding remarks}
In this article, we construct stationary exact solutions to Einstein-Gauss-Bonnet gravity based on four-dimensional self-dual Atiyah-Hitchin geometry. We find two different solutions for the metric function which are exact solutions to the five-dimensional Einstein-Gauss-Bonnet field equations. To the best of our knowledge, these solutions are the first known solutions to five-dimensional Einstein-Gauss-Bonnet theory where the base space is the self-dual Atiyah-Hitchin space.  we notice that the metric function is regular everywhere in spacetime.  
To verify that the solutions indeed satisfy in all the field equations, we consider some numerical calculation, as the analytical field equations are so long that is almost impossible to verify that they satisfy analytically the field equations. 

We conclude with a few comments about extending the solutions in this article. Though we consider the dependence of the five-dimensional metric function on only one coordinate, we may find other numerical solutions, where the metric function depends on more coordinates. 
We are also interested in finding the solutions to Einstein-Gauss-Bonnet theory in presence of the cosmological constant.  The cosmological solutions with the Atiyah-Hitchin space as a part of bulk spacetime, can be studied in the context of (A)dS/CFT correspondence \cite{last1}-\cite{last4}.  We also leave the study of the thermodynamics of the solutions for a forthcoming article.

\bigskip
\bigskip
\bigskip

{\Large Acknowledgements}

This work was supported by the Natural Sciences and Engineering Research Council of Canada.

\bigskip
\bigskip
\bigskip

\section{Appendix A} \label{A1}
The coefficients of quartic equation (\ref{QUA}), are given by the following expressions
\begin{eqnarray}
d_1&=&2{ a_1}{ b_1}{ c_1}-2{ a_1}{ b_3}{ c_1}-2{ a_1}{ c_2}+2{ a_1}{ c_4},\\
d_2&=& 2{ a_1}{ b_1}{ c_1}-2{ a_1}{ b_4}{ c_1}-{ a_2}{ b_1}{ c_1}-2{ b_1}{ b_3}{ c_1}-2{ a_1}{ c_3}+2{ a_1}{ c_5}+{ c_2}{ a_2}+{ c_4}{ b_1}+{ c_2}{ b_3},\\
d_3&=& -2{ a_1}{ b_5}{ c_1}-{ a_2}{ b_1}{ c_1}-{ a_3}{ b_1}{ c_1}-{ b_1}{ b_3}{ c_1}-2{ b_1}{ b_4}{ c_1}-{ b_2}{ b_3}{ c_1}+2{ c_6}{ a_1}+{ c_3}{ a_2}+{ c_2}{ a_3}\nonumber \\
&+&{ c_5}{ b_1}+{ c_4}{ b_2}+{ c_3}{ b_3}+{ c_2}{ b_4},\\
d_4&=& -{ a_3}{ b_1}{ c_1}-{ a_4}{ b_1}{ c_1}-{ b_1}{ b_4}{ c_1}-2{ b_1}{ b_5}{ c_1}-{ b_2}{ b_4}{ c_1}+{ c_3}{ a_3}+{ c_2}{ a_4}+{ c_6}{ b_1}+{ c_5}{ b_2}\nonumber\\
&+&{ c_3}{ b_4}+{ c_2}{ b_5},\\
d_5&=&-{ a_4}{ b_1}{ c_1}-{ b_1}{ b_5}{ c_1}-{ b_2}{ b_5}{ c_1}+{ c_3}{ a_4}+{ c_6}{ b_2}+{ c_3}{ b_5},
\end{eqnarray}
where $a_i,b_i,i=1,\cdots ,5$ and $c_i,\,i=1,\cdots ,6$ are given in terms of Atiyah-Hitchin metric functions (\ref{a-th})-(\ref{c-th}) by
\begin{eqnarray}
a_1&=&3,\\
a_2&=&12~\big (a^2-2( b+ c) a+(b-c)^2\big), \\
a_3&=&-32~(-c+a+b) (-b+c+a) (c+a+b) (a-b-c),\\
a_4&=&-512~ \big(a^6+(-b-c) a^5+(-b^2+3 b c-c^2)a^4+2 (c+b) (b-c)^2 a^3\nonumber\\
      &-&(b^2+4 b c+c^2) (b-c)^2 a^2-(c+b) (b-c)^4 a+(b^2-b c+c^2) (b-c)^2 (c+b)^2\big),\nonumber\\
      &&\\
a_5&=&24 ~a^2 b^2 c^2,
\end{eqnarray}
\begin{eqnarray}
b_1&=&-32~c\big(a+b-c\big),\\
b_2&=&-64~\big(-2 c^4+(a+b) c^3+(a-b)^2 c^2-(a+b) (a-b)^2 c+(a-b)^2 (a+b)^2\big),\nonumber\\
&&\\
b_3&=&-12~\big(5 c^2-6 ( a+ b) c+(a-b)^2\big),\\
b_4&=&160~\big(-(13/5)c^4+2(a+b) c^3-(12/5) a b c^2\nonumber\\
&-&2/5 (a+b) (a-b)^2 c+(a-b)^2 (a+b)^2\big),\\
b_5&=&512~\Big(-c^6+2(a+b) c^5+(-a^2-a b-b^2) c^4+(a^2+b^2) (a-b)^2 c^2\nonumber\\
      &+&(-2 a^5+2 a^4 b+2 a b^4-2 b^5) c+(a^2-a b+b^2) (a-b)^2 (a+b)^2\Big),
      \end{eqnarray}
      
and

 \begin{eqnarray}
c_1&=&a+b,\\
c_2&=&32~\Big(a^3-a^2 c-a b c+b^2 (b-c)\Big),\\
c_3&=&64~\Big(2 a^5+(b-c)a^4+(b^2+bc-c^2)a^3+(b^3-2bc^2+c^3)a^2+(b-c)(b^3+2b^2c+c^3)a\nonumber\\
&+&(b-c)b(c+b)(2b^2-bc+c^2)\Big),\\
c_4&=&-12~\big(5a^3-(b+6 c)a^2+(-b^2-8 bc+c^2)a+(5b-c)(b-c)b\big),\\
c_5&=&-32~\Big(13a^5+(3b-10c)a^4+2(4b+ c)b a^3+2(4b^3-bc^2+c^3)a^2\nonumber
\\&+&(5 c+3b)(b-c)(b^2+c^2)a+13 b^5-10 b^4c+2b^2 c^3-5bc^4\Big),\\     
c_6&=&-512~\Big(a^7+(-b-2c)a^6+(b^2-bc+c^2)a^5+(-b^3-2b^2c+bc^2)a^4-(b^2+c^2)(b-c)^2a^3\nonumber
\\&+&(b-c)(b^4-b^3c-3b^2c^2+bc^3-2c^4)a^2-(b-c)(b^5+2b^4c+b^3c^2-b^2c^3+2bc^4-c^5)a\nonumber\\
&+&b(c+b)(b^2+c^2)(b-c)^3\Big).
\end{eqnarray}

\section{Appendix B} \label{A2}

The four solutions to equation (\ref{QUA}) are given by

\begin{eqnarray}
x_1&=&-\frac{{d_2}}{4 {d_1}}-\frac{1}{2} e_1^{1/2}-\frac{1}{2} (e_2-e_3)^{1/2}, \\
x_2&=&-\frac{{d_2}}{4 {d_1}}-\frac{1}{2} e_1^{1/2}+\frac{1}{2} (e_2-e_3)^{1/2},   \\
x_3&=&-\frac{{d_2}}{4 {d_1}}+\frac{1}{2} e_1^{1/2}-\frac{1}{2}  (e_2+e_3)^{1/2}, \\
x_4&=&-\frac{{d_2}}{4 {d_1}}+\frac{1}{2} e_1^{1/2}+\frac{1}{2}  (e_2-e_3) ^{1/2},
\end{eqnarray}
where
\begin{eqnarray}
e_1&=&{\frac{{d_2}^2}{4{d_1}^2}-\frac{2 {d_3}}{3 {d_1}}+\frac{D}{3\sqrt[3]{2} {d_1}}+\frac{\sqrt[3]{2} \left(12 {d_1}{d_5}-3 {d_2} {d_4}+{d_3}^2\right)}{3 {d_1}D}},\label{e1}\\
e_2&=&\frac{{d_2}^2}{2{d_1}^2}-\frac{4 {d_3}}{3{d_1}}-\frac{D}{3\sqrt[3]{2} {d_1}}-\frac{\sqrt[3]{2} \left(12 {d_1}{d_5}-3 {d_2} {d_4}+{d_3}^2\right)}{3 {d_1}D},\label{e2}\\
e_3&=&\frac{-\frac{{d_2}^3}{{d_1}^3}+\frac{4 {d_2}{d_3}}{{d_1}^2}-\frac{8 {d_4}}{{d_1}}}{4\sqrt{\frac{{d_2}^2}{4 {d_1}^2}-\frac{2 {d_3}}{3{d_1}}+\frac{D}{3 \sqrt[3]{2} {d_1}}+\frac{\sqrt[3]{2}\left(12 {d_1} {d_5}-3 {d_2}{d_4}+{d_3}^2\right)}{3 {d_1} D}}}.\label{e3}
\end{eqnarray}
The function $D$ in equations (\ref{e1})-(\ref{e3}) is given by
\begin{eqnarray*}
D&=&\Bigg(\sqrt{\left(-72 {d_1} {d_3}{d_5}+27 {d_1} {d_4}^2+27 {d_2}^2 {d_5}-9{d_2} {d_3} {d_4}+2 {d_3}^3\right)^2-4 \left(12{d_1} {d_5}-3 {d_2} {d_4}+{d_3}^2\right)^3}
\\&-&72{d_1} {d_3} {d_5}+27 {d_1} {d_4}^2+27 {d_2}^2{d_5}-9 {d_2} {d_3} {d_4}+2 {d_3}^3~~\Bigg)^{\frac13}.
\end{eqnarray*}

\end{document}